\documentclass[twocolumn,superscriptaddress,showpacs,aps,amsmath,amssymb,pra]{revtex4}

\newif\ifFigureFolder
\newif\ifNotArXiv

\FigureFolderfalse	
\NotArXivfalse		

\usepackage{subfigure}
\usepackage{graphicx}
\usepackage{color}

\ifNotArXiv
	\usepackage{todonotes}
	\usepackage{hyperref}
\fi
\ifFigureFolder
	\graphicspath{{./figures/}}
\fi

\begin{document}

\title{Creation of a squeezed photon distribution using artificial atoms with broken inversion symmetry}

\author{Martin Koppenh\"ofer}
\affiliation{Institut f\"ur Theoretische Festk\"orperphysik, Karlsruhe Institute of Technology, D-76128 Karlsruhe, Germany}

\author{Michael Marthaler}
\affiliation{Institut f\"ur Theoretische Festk\"orperphysik, Karlsruhe Institute of Technology, D-76128 Karlsruhe, Germany}

\pacs{42.50.Pq,42.50.Dv,81.07.Ta,74.78.Na}

\date{\today}

\begin{abstract}
We consider a two level system with both a transversal and a longitudinal coupling to the electromagnetic field of a resonator. 
Using a polaron transformation, this Hamiltonian can be mapped onto a Jaynes-Cummings Hamiltonian with generalized field operators 
acting on the electromagnetic field in the resonator. 
In contrast to the usual ladder operators $a$ and $a^\dagger$, these operators exhibit a non-monotous coupling 
strength with respect to the number $n$ of photons in the resonator. Especially, there are roots of the coupling between qubit and resonator at certain photon numbers $n_0$. We show that this effect can be exploited to generate 
photon-number squeezed light, characterized by a Fano factor $F \ll 1$, with a large number of photons (e.g.\ of the order of $10^4$).
\end{abstract}

\maketitle

\newcommand{\ket}[1]{\left\vert #1 \right\rangle}
\newcommand{\bra}[1]{\left\langle #1 \right\vert}
\newcommand{\abs}[1]{\left\vert #1 \right\vert}
\newcommand{\erw}[1]{\left\langle #1 \right\rangle}

\section{Introduction}
Lasers serve as the source of coherent light in spectroscopical and interferometric measurements. The precision of these measurements is fundamentally limited due to shot noise caused by the quantized nature of light and the photon statistics of the radiation source. To circumvent this limitation, squeezed light has been theoretically proposed \cite{Caves-PhysRevD.23.1693,Scully-PhysRevA.53.467,Agarwal-OpticsLetters.38.14.2563,Didier-1502.00607,Berchera-1501.07516,Knott-PhysRevA.89.53812} and successfully applied in physical \cite{Polyik-PhysRevLett.68.3020,Iwasawa-PhysRevLett.111.163602,Miller-PhysRevD.91.062005} and biological experiments \cite{Taylor-NatPhot2012-346,Whittaker-arXiv}. 

To create squeezed light, non-linear processes are necessary. In the optical regime, squeezed light is created using a conventional laser as input source for a non-linear optical material that exploits higher order processes like wave mixing
or parametric down conversion 
\cite{Shelby-PRL57-691,Dong-OptLett33-116-2008,Mehmet-PRA81-013814-2010} to create squeezed light. In the microwave regime, superconducting parametric devices and non-linear oscillators have been demonstrated and are beeing used for parametric \cite{Devoret_Siddiqi_Quantum_Review,Devoret_Parametric} and bifurcation amplification \cite{Nonlinear_oscillator}.
It is even possible to build non-linear  superconducting oscillators in the quantum regime 
\cite{nonlinear_oscillator_Jochen,Driven_Quantum_Spectrum,Driven_Quantum_Tunneling}. However, while parametric processes can be used to generate quadrature 
squeezing, in this work we will study the creation of a squeezed photon distribution \cite{KnightTutorialQuantumOptics}. 

A laser uses atoms as active medium to create photons. For natural atoms, all non-diagonal matrix elements of the dipole coupling between cavity and atom vanish 
because of the inversion symmetry of the atomic Coulomb potential. In terms of a Jaynes-Cummings model, this means that there is a pure $\sigma_x$-coupling to the radiation field.
However, every setup that breaks inversion symmetry will exhibit an additional $\sigma_z$-coupling to the radiation field. As we will show below, this gives rise to photon-number squeezing already in leading order. 
Such $\sigma_z$-couplings exist, for instance, in superconducting circuits \cite{Metamaterial_Pascal}, quantum dots \cite{Childress-PRA-69-042302} and molecules \cite{Meath-MolPhys51-3-585}. It has been shown that a quantum dot with broken inversion symmetry in a microcavity acts as a non-linear optical element \cite{SavenkoAsymmetricQuantumDot}. Lasing with organic molecules is a very applied research field \cite{MolecularLaser} and lasing devices based on solid state qubits have been studied \cite{Armour-SSET-Micromaser,Doube_Dot_Jinshuang}. For superconducting devices, non-linearities based on the Josephson effect have been proposed as a way to create 
squeezed photon distributions \cite{Marthaler-PhysRevLett.101.147001,Marthaler_Squeezed,Squeezed_Biased_Junction}. In experimentally realized lasing devices using superconducting qubits \cite{Astafiev-Nat-449-588,Chen-PhysRevB.90.020506,Dressed_State_Lasing_Jena}
or gate defined double dots \cite{Doube_dot_Laser_Petta},
the $\sigma_z$-coupling between artifical atom and cavity field is present but has not yet been studied. In addition, the average number of photons in the laser cavity is quite low (e.g.\ less than $200$ in Ref. \onlinecite{Chen-PhysRevB.90.020506}). 

\begin{figure}
	\centering
	\ifFigureFolder
		\resizebox{!}{.285\textwidth}{\Large \input{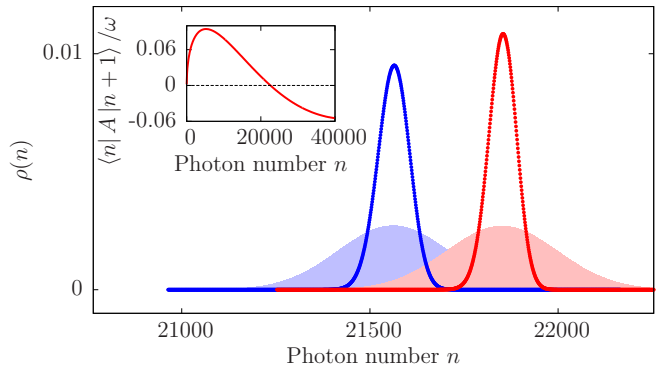}}
	\else 
		\resizebox{!}{.285\textwidth}{\Large \input{Fig1}}
	\fi
	\caption{(Color online) Photon statistics $\rho(n)$ in the polaron frame for $g=.0067$, $\theta=\pi/10$, $\Gamma_\uparrow=.006$, $\Gamma_\downarrow=.0001$, $\Gamma_\varphi^*=.001$, $\Delta = 0$. Red dots: One atom, $\kappa=1 \cdot 10^{-7}$. Blue dots: $100$ atoms, $\kappa = 1 \cdot 10^{-5}$, $S_z^k = 25.8$. All rates and couplings are given in units of $\omega$. The distributions correspond to a Fano factor $F_0=.0621$ and $F_0=.0819$ in the polaron frame, respectively. The Fano factor in the photon-number frame, $F$, is enhanced by corrections due to the polaron transformation, yielding $F = 0.0622$ and $F=0.1442$ (strongly squeezed light), respectively. The  average photon number is $\erw{n}=21851$ and $\erw{n}=21562$, respectively. The filled faint red and blue curves represent the photon statistics of an ordinary laser producing classical light with the same average photon number $\erw{n}$. Inset: Coupling matrix element $\bra{n} A \ket{n+1}$. The root at the photon number $n_0=22599$ causes the squeezed photon states.}
	\label{fig:PhotonStatisticsAndMatrixElement}
\end{figure}

In this paper, we examine a system described by a general Hamiltonian including both a $\sigma_x$- and a $\sigma_z$-coupling between atom and radiation field. 
Using a polaron transformation, the general Hamiltonian can be mapped onto a Jaynes-Cummings Hamiltonian with generalized field operators that exhibit a non-monotonous coupling strength with respect to the number $n$ of photons in the resonator (see inset in Fig.~\ref{fig:PhotonStatisticsAndMatrixElement}). If population inversion is established, the photon number in the cav\-ity starts to increase. The stationary average photon number in the laser cavity is given by the balance of photon creation and photon loss rates in the system. Near $n_0$, the position of the root of the generalized field operator, the photon creation process breaks down, because of the decreasing coupling between atom and resonator. This establishes a squeezed photon distribution with an average photon number of the order of $n_0$. 
The average photon number 
can be very large, e.g.\ of the order of $10^4$, for realistic parameters. Simultaneously, a strong squeezing, characterized by a Fano factor of $F \ll 1$, can be reached (see Fig.~\ref{fig:PhotonStatisticsAndMatrixElement}). 

\section{Hamiltonian}
\label{sec:hamiltonian}
The system under consideration is described by a Hamiltonian consisting of an artificial atom modelled by a two level system, interacting with the quantized electromagnetic field of a resonator, 
\begin{align}
	H = \frac{1}{2} \epsilon \sigma_z + \hbar \omega a^\dagger a + \hbar g \left( \cos(\theta) \sigma_z + \sin(\theta) \sigma_x \right) \left( a + a^\dagger \right).
	\label{eqn:Hamiltonian1Atom}
\end{align} 
A generalization to an arbitrary number of atoms follows below. The operators 
$\sigma_i$ with $i \in \{x,y,z\}$ denote the Pauli-matrices, $a$  ($a^\dagger$) is the annihilation (creation) operator of a photon with frequency $\omega$ and $\epsilon$ is the level splitting energy of the two level system. The photon field and the two level system are coupled with coupling strength $g$. In constrast to the standard Jaynes-Cummings model, our system has both a transversal and a longitudinal coupling to the 
electromagnetic field. The relative coupling strength is characterized by the angle $\theta$.

Eq.~\ref{eqn:Hamiltonian1Atom} describes an effective lasing Hamiltonian, where we did not explicitly consider the third state which we need to establish population inversion. 
The pumping process will be modelled by a Lindblad-term in the master-equation of this system and will be described below. It contains the effective pumping rates between the upper and the lower lasing state.

The Hamiltonian~\eqref{eqn:Hamiltonian1Atom} can be mapped onto the well-known Jaynes-Cummings Hamiltonian using the polaron transformation
\begin{align*}
	U = \exp \left[ i p \sigma_z \right] = \exp \left[ \frac{g}{\omega} \cos(\theta) \left( a - a^\dagger \right) \sigma_z \right] ~.
\end{align*} 
For convenience, we introduce the operators 
\begin{align*}
	x &= x_0 \left( a^\dagger + a\right) = \hbar g \sin (\theta) \left( a^\dagger + a \right) ~,\\
	p &= i p_0 \left( a^\dagger - a \right) = i \frac{g}{\omega} \cos (\theta) \left( a^\dagger - a \right)~.
\end{align*} 
The transformation yields
\begin{align}
	H_\mathrm{p} 
	= U^\dagger H U 
	= \frac{1}{2} \epsilon \sigma_z + \hbar \omega a^\dagger a + \left( \sigma_+ A + \sigma_- A^\dagger \right) 
	\label{eqn:HamiltonAfterPolaron}
\end{align} 
with the operators $A = e^{- i p} x e^{- i p}$ and $A^\dagger = e^{i p} x e^{i p}$ instead of the pure annihilation and creation operators known from the Jaynes-Cummings Hamiltonian.

As basis we choose the direct product of the resonator states, $a^{\dag}a \ket{n} = n \ket{n}$, and the states of the two level system $\sigma_z \ket{\uparrow,\downarrow}=\pm \ket{\uparrow,\downarrow}$. The state $\ket{n}$ is defined in the polaron frame, $\ket{n}
\ket{\sigma} = U^\dagger \ket{n}_\mathrm{c} \ket{\sigma}$, where $\ket{n}_\mathrm{c}$ is the state with exactly $n$ photons in the resonator cavity. The matrix elements of $A$ ($A^\dagger$) are purely real and can be expressed in terms of the generalized Laguerre polynomials, 
\begin{align}
	\bra{n} A^{(\dagger)} \ket{n+m} &= \pm \frac{m}{2} \frac{x_0}{p_0} T^\pm_{n,m}~,
	\label{eqn:MatrixElement}
\end{align} 
with $A^\dagger$ taking the upper sign and $n \in \mathbb{N}$, $m \in \mathbb{Z}$, $m \geq -n$. For $0 < m < \infty$, $T^\pm_{n,m}$ is given by
\begin{align*}
	T^\pm_{n,m} 
	&= \bra{n} e^{\pm 2 i p} \ket{n+m} \\
	&= (\pm 1)^m e^{-2 p_0^2} \left( 2 p_0 \right)^m \sqrt{\frac{n!}{(n+m)!}} L_n^m (4 p_0^2)~,
\end{align*} 
where $L_n^m(x)$ are the associated Laguerre polynomials. For $-n \leq m < 0$, one finds
\begin{align*}
	T^\pm_{n,m} 
	&= \bra{n} e^{\pm 2 i p} \ket{n+m} \\
	&= (\mp 1)^{\abs{m}} e^{-2 p_0^2} \left( 2 p_0 \right)^{\abs{m}} \sqrt{\frac{(n-
\abs{m})!}{n!}} L_{n-\abs{m}}^{\abs{m}} (4 p_0^2)~.
\end{align*} 

For now, we focus only on transitions that are almost energy-conserving. This step will be justified below. Given this assumption, the coupling between atom and resonator depends only on the matrix element $\bra{n} A \ket{n+1}$, i.e. $m = 1$. Choosing $\theta=\pi/2$ reproduces the well known $\sqrt{n}$ behaviour of the Jaynes-Cummings model. However, for $\theta < \pi/2$ the matrix element shows a non-monotonous dependence of the number $n$ of photons in the resonator (inset in Fig.~\ref{fig:PhotonStatisticsAndMatrixElement}). Especially, there are photon numbers $n_0$ where the matrix element is close to zero. There, the atom isn't able to further increase the number of photons in the resonator. As discussed in the next section, this is accompanied by a squeezed photon distribution. 

\section{Photon statistics}
We calculate the photon statistics of the laser analogously to Ref.~\onlinecite{Scully-Chap11}: The system is described by the master-equation for its density matrix $\rho$,
\begin{align}
	\dot{\rho} &= - \frac{i}{\hbar} \left[ H_\mathrm{p}, \rho \right] + L_\mathrm{R} \rho + L_\mathrm{Q} \rho~.
	\label{eqn:MasterEquationWholeSystem}
\end{align} 
The Lindblad-superoperators are given by
\begin{align*}
	L_\mathrm{R} \rho 
	&= \frac{\kappa}{2} \left( 2 a \rho a^\dagger - a^\dagger a \rho - \rho a^\dagger a \right) ~,\\
	L_\mathrm{Q} \rho 
	&= \frac{\Gamma_\downarrow}{2} \left( 2 \sigma_- \rho \sigma_+ - \rho \sigma_+ \sigma_- - \sigma_+ \sigma_- \rho \right) \\
	&+ \frac{\Gamma_\uparrow}{2} \left( 2 \sigma_+ \rho \sigma_- - \rho \sigma_- \sigma_+ - \sigma_- \sigma_+ \rho \right) \\
	&+ \frac{\Gamma_\varphi^*}{2} \left( \sigma_z \rho \sigma_z - \rho \right)~,
\end{align*} 
where $\kappa$ is the damping rate of the resonator, $\Gamma_\uparrow$ and $\Gamma_\downarrow$ are the effective pumping rates between the lasing states (including relaxation effects) and $\Gamma_\varphi^*$ is the pure dephasing rate of the atom. We will show below that this form of the Lindblad-superoperators is a good approximation even after the polaron transformation. 

We derive an effective equation of motion for the reduced density matrix of the resonator,
\begin{align*}
	\rho_\mathrm{r} = \mathrm{Tr}_\mathrm{atom} (\rho)~,
\end{align*}
where the trace is only taken over the atomic states.
Tracing out the atomic states in Eq.~\ref{eqn:MasterEquationWholeSystem} yields
\begin{align}
	\dot{\rho}_\mathrm{r} = &- \frac{i}{\hbar} \left[ \hbar \omega a^\dagger a , \rho_\mathrm{r} \right] \nonumber \\
	&- \frac{i}{\hbar} \mathrm{Tr}_\mathrm{atom} \left[ (\sigma_+ A + \sigma_- A^\dagger ) , \rho \right] + L_\mathrm{R} \rho_\mathrm{r} ~.
	\label{eqn:MasterEquationReducedResonator}
\end{align}

To evaluate the second term, we need to solve the remaining equation of motion for $\rho$, 
\begin{align*}
	\dot{\rho} = - \frac{i}{\hbar} \left[ H_\mathrm{p} , \rho \right] + L_\mathrm{Q} \rho~.
\end{align*}
We write this as a system of four coupled differential equations for the matrix elements of all possible combinations of atomic states.
The matrix elements are denoted by $\rho_{r\,p,s\,q} = \bra{p} \bra{r} \rho \ket{s} \ket{q}$ with $r,s \in \{ \uparrow,\downarrow\}$ and $p,q \in \mathbb{N}_0$.
If $\rho_{\downarrow p,\downarrow q}$ and $\rho_{\uparrow p+1, \uparrow q+1}$ are eliminated using $(\rho_\mathrm{r})_{p,q} = \rho_{\uparrow p,\uparrow q} + \rho_{\downarrow p,\downarrow q}$, the system can be cast into the form
\begin{align*}
	\dot{\vec{R}} = M \cdot \vec{R} + \vec{A}(\rho_\mathrm{r})
\end{align*}
with 
\begin{align*}
	\vec{R} = \begin{pmatrix}
		\rho_{\uparrow p,\uparrow q} \\
		\rho_{\uparrow p,\downarrow q+1} \\
		\rho_{\downarrow p+1,\uparrow q} \\
		\rho_{\downarrow p+1,\downarrow q+1} 
	\end{pmatrix}~.
\end{align*}
$M$ and $\vec{A}(\rho_\mathrm{r})$ are given in the appendix. 

The reduced density matrix $\rho_\mathrm{r}$ of the resonator evolves much more slowly than the density matrix $\rho$ of the whole system. Therefore, we can use an adiabatic approximation and take the stationary solution
\begin{align*}
	\vec{R} = - M^{-1} \vec{A}~.
\end{align*}

Now, the trace in Eq.~\ref{eqn:MasterEquationReducedResonator} can be evaluated and $\rho_\mathrm{r}$ can be calculated. Its diagonal entries 
\begin{align*}
	\rho(n) = \bra{n} \rho_\mathrm{r} \ket{n}
\end{align*}
are the probability distribution of the photon number states in the polaron frame and obey the recursion relation
\begin{align}
	\rho(n) &= f(n) \cdot \rho(n-1) ~,
	\label{eqn:PhotonStatisticsRecursionRelation} \\
	f(n) &= \frac{\xi(n-1) \cdot \Gamma_\uparrow}{\kappa \cdot n + \xi(n-1) \cdot \Gamma_\downarrow} \nonumber ~.
\end{align} 
The parameters are definied as follows:
\begin{align}
	\xi(n-1) &= \frac{\mathcal{A}}{1 + \mathcal{B} \cdot N(n-1)} \frac{\abs{\bra{n-1} A \ket{n}}^2}{\hbar^2 g^2} \nonumber \\
	N(n) &= \frac{\Delta^2}{4 g^2} \frac{\Gamma_1}{\frac{\Gamma_1}{2} + \Gamma_\varphi^*} + \frac{\abs{\bra{n} A \ket{n+1}}^2}{\hbar^2 g^2} 
	\label{eqn:NofP} \\
	\mathcal{A} &= \frac{2 g^2}{\Gamma_1 \left( \frac{\Gamma_1}{2} + \Gamma_\varphi^* \right)} \nonumber \\
	\mathcal{B} &= \frac{4 g^2}{\Gamma_1 \left( \frac{\Gamma_1}{2} + \Gamma_\varphi^* \right)} \nonumber \\
	\Delta &= \frac{\epsilon}{\hbar} - \omega \nonumber \\
	\Gamma_1 &= \Gamma_\uparrow + \Gamma_\downarrow \nonumber
\end{align} 

\begin{figure}
	\centering
	\ifFigureFolder
		\resizebox{!}{.285\textwidth}{\Large \input{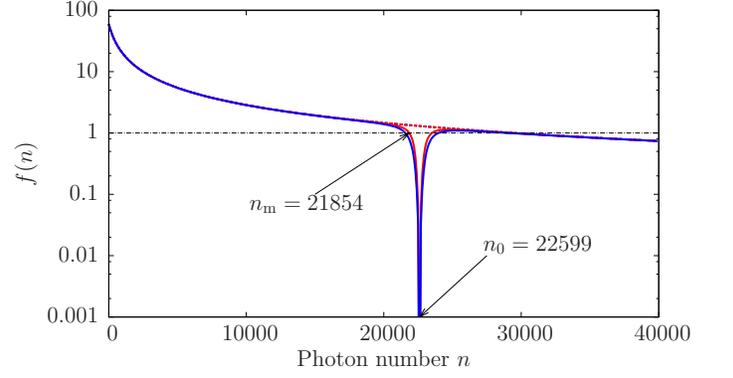}}
	\else 
		\resizebox{!}{.285\textwidth}{\Large \input{Fig2}}
	\fi
	\caption{(Color online) Recursion coefficient $f(n)$ of the photon statistics $\rho(n) = f(n) \cdot \rho(n-1)$ for one atom (red) and $100$ atoms (blue). The maximum of $\rho(n)$ is situated at $n_\mathrm{m}$ defined by $f(n_\mathrm{m})=1$ and $f'(n_\mathrm{m}) < 0$. A conventional laser (pure $\sigma_x$-coupling, dashed lines) will give $n_\mathrm{m} = 29450$. In case of both $\sigma_x$- and $\sigma_z$-coupling (solid lines), the root of $\bra{n} A \ket{n+1}$ at $n_0 = 22599$ gives a maximum at much smaller $n_\mathrm{m} = 21854$ and $n_\mathrm{m} = 21565$, respectively. 
	As the slope of the solid lines at $n_\mathrm{m}$ is sharp, a squeezed state is created with $F \ll 1$.
	Plot parameters are identical to Fig.~\ref{fig:PhotonStatisticsAndMatrixElement}.}
	\label{fig:RecursionCoefficient}
\end{figure}

The photon distribution $\rho(n)$ has a local maximum at photon numbers $n_\mathrm{m}
$ with $f(n_\mathrm{m}) = 1$ and $f'(n_\mathrm{m}) < 0$. Fig.~\ref{fig:RecursionCoefficient} compares $f(n)$ for pure $\sigma_x$-coupling ($\theta = \pi/2$, dashed lines) and generalized couplings (solid lines). 

For $\theta = \pi/2$, the recursion relation~\eqref{eqn:PhotonStatisticsRecursionRelation} can be solved analytically. Far above the lasing threshold, it is a Poissonian distribution \cite{Scully-Chap11}. As $f(n)$ decreases monotonically for $\theta = \pi/2$, there is only one maximum of $\rho(n)$. 

For $\theta \neq \pi/2$, $f(n)$ has a root if $\bra{n} A \ket{n+1} = 0$. In general, there are now several $n_\mathrm{m}$ fulfilling the criteria for a local maximum, situated at much smaller photon numbers than the average photon number for $\theta = \pi/2$. The absolute value of the slope $\abs{f'(n_\mathrm{m})}$ at these photon numbers is larger than in the case of $\theta = \pi/2$. As discussed in the next section, this yields a photon-number squeezed state. 

In principle, there could be several local maxima of $\rho(n)$, but in general only one of these maxima has a probability of the order of unity, unless the lasing parameters are carefully tuned. 

\section{Fano factor}
We measure the squeezedness of the radiation using the Fano factor $F$ defined by
\begin{align*}
	F &= \frac{\erw{n^2}_\mathrm{c} - \erw{n}_\mathrm{c}^2}{\erw{n}_\mathrm{c}} \geq 0~.
\end{align*} 
with $n= a^\dagger a$. As introduced above, $\ket{n}_\mathrm{c}$ denotes the state with $n$ photons in the resonator cavity. The Fano factor is $F=1$ if $\rho(n)$ is Poissonian, $F = \erw{n} + 1$ if $\rho(n)$ describes a thermal state and $F < 1$ if $\rho(n)$ describes a photon-number squeezed state. 

Expressed by states in the polaron frame, the Fano factor $F$ is 
\begin{align*}
	F &= \frac{\erw{n^2 + 2 p_0^2 n} - \erw{n}^2 + p_0^2}{\erw{n} + p_0^2} \\
	&+ \frac{p_0^2 \sum_{n=0}^\infty \left( \sqrt{n} \sqrt{n-1} \rho_{n,n-2} + \sqrt{n+1} \sqrt{n+2} \rho_{n,n+2} \right)}{\erw{n} + p_0^2} ~.
\end{align*} 
As $\rho_{n,n \pm 2} \approx 0$ (there are no correlations of different photon-number states) and $p_0^2 = \mathcal{O}\left((\frac{g}{\omega})^2 \right) \ll \erw{n}$, we get
\begin{align*}
	F \approx F_0 + 2 p_0^2~,
\end{align*} 
where $F_0 = \frac{\erw{n}^2 - \erw{n}^2}{\erw{n}}$ is now defined in the polaron frame. 
Given an arbitrary $\rho(n)$, $F_0$ can be calculated numerically. We will show that the value of $F_0$ depends on the (negative valued) slope of $f(n)$ at $f(n_\mathrm{m}) = 1$. For that purpose, we linearize $f(n)$ around the maximum photon number $n_\mathrm{m}$, defined by $f(n_\mathrm{m}) = 1$ and $f'(n_\mathrm{m}) = - c$, $c > 0$, 
\begin{align*}
	f(n) \approx 1 - c \left( n - n_\mathrm{m} \right)~.
\end{align*} 
This approximation is exact near $n_\mathrm{m}$. As $\rho(n)$ drops fast around $n_\mathrm{m}$, deviations from the linearized formula are only large in a region where $\rho(n) \ll 1$. These regions don't contribute to the calculation of the Fano factor. Of course, the approximation can only be used for $n < n_\mathrm{m} + \frac{1}{c}$, as $f(n)$ becomes negative for larger $n$. A calculation of $F_0$ using the linearized $f(n)$ yields 
\begin{align*}
	F_0 = \frac{1}{c} \frac{1}{\erw{n}} = \frac{1}{c} \frac{1}{n_\mathrm{m}-1}~.
\end{align*} 
which is valid as long as $c \gg e^{-n_\mathrm{m}}$.
The steeper $f(n)$ at $n_\mathrm{m}$, the smaller the Fano factor. By tuning the lasing parameters in such a way, that $f(n) = 1$ is fulfilled in one of the regions near a root of $\bra{n} A \ket{n+1}$ and that $f(n)$ exhibits a sharp slope there, one achieves values $F \ll 1$.  

\section{Multi-atom lasing}
The number of photons in the resonator can be increased by taking $M$ artificial atoms (with $M > 1$). Therefore, we generalize our model to $M$ identical atoms coupled to a common resonator,
\begin{align*}
	H &= \sum_{i=1}^M \frac{1}{2} \epsilon \sigma_z^i + \hbar \omega a^\dagger a \\
	&+ \hbar g \sum_{i=1}^M \left( \cos(\theta) \sigma_z^i + \sin(\theta) \sigma_x^i \right) \left( a + a^\dagger \right)~.
\end{align*} 
The superscript $i$ of the Pauli-matrices denotes the atom they act on. The polaron transformation is generalized as well,
\begin{align}
	U = \exp \left[ i p \sum_{j=1}^M \sigma_z^j \right]~,
	\label{eqn:PolaronMAtoms}
\end{align} 
$p$ is defined as above. Transforming $H$ yields
\begin{align*}
	H_\mathrm{p} 
	&= U^\dagger H U \\
	&= \sum_{i=1}^M \frac{1}{2} \epsilon \sigma_z^i + \hbar \omega a^\dagger a + \sum_{i=1}^M \left( \sigma_+^i A + \sigma_-^i A^\dagger \right) \\
	&- 2 x_0 p_0 \sum_{i \neq j = 1}^M \left( \sigma_+^i \sigma_z^j e^{-2 i p} + \sigma_-^i \sigma_z^j e^{2 i p} \right) \\
	&- \hbar \omega p_0^2 M - \hbar \omega p_0^2 \sum_{i \neq j = 1}^M \sigma_z^i \sigma_z^j~.
\end{align*} 
$A$ and $A^\dagger$ are defined as above. The last term introduces correlations between all atoms, the $\sigma_\pm^i \sigma_z^j$ terms introduce photon-number dependent couplings between atoms via the $e^{\pm 2 i p}$ terms. To solve this, we perform a mean-field-approximation,
\begin{align*}
	\sigma_z^l \sigma_z^j &\approx \sigma_z^l \erw{\sigma_z^j} + \erw{\sigma_z^l} \sigma_z^j - \erw{\sigma_z^l} \erw{\sigma_z^j} ~, \\
	\sigma_\pm^i \sigma_z^j &\approx \sigma_\pm^i \erw{\sigma_z^j} + \erw{\sigma_\pm^i} \sigma_z^j - \erw{\sigma_\pm^i} \erw{\sigma_z^j} = \sigma_\pm^i \erw{\sigma_z^j}~,
\end{align*} 
where in the last step we assumed that only energy-conserving matrix elements of $\rho$ are finite, implying $\erw{\sigma_\pm^i} = 0$. Defining
\begin{align*}
	S_z^j = \sum_{\stackrel{i \neq j}{i=1}}^M \erw{\sigma_z^i}
\end{align*} 
and assuming that all atoms are identical, we can map the $M$-atom Hamiltonian on the single-atom case~\eqref{eqn:HamiltonAfterPolaron} with modified level splitting energy $\epsilon \to E(S_z^k) = \epsilon - 4 \hbar \omega p_0^2 S_z^k$, modified field operators $A \to A(S_z^k) = e^{- i p} x e^{- i p} - 2 x_0 p_0 S_z^k e^{-2 i p}$ and an irrelevant constant term. $k$ is the index of an arbitrarily chosen atom. Note that both terms in $A(S_z^k)$ are proportional to $e^{- 2 i p}$, so that the roots of the coupling matrix elements are not changed, but the steepness of the recursion coefficient $f(p)$ changes (Fig.~\ref{fig:RecursionCoefficient}). 

Calculating the photon statistics, there is a change in $\xi(n-1)$ due to the increased number of atoms and the modified field operators,
\begin{align*}
	\xi(n-1) &= M \cdot \frac{\mathcal{A}}{1 + \mathcal{B} \cdot N(n-1)} \frac{\abs{\bra{n-1} A(S_z^k) \ket{n}}^2}{(\hbar g)^2} ~,\\
	\Delta(S_z^k) &= \Delta - 4 \omega p_0^2 S_z^k~.
\end{align*} 
$\mathcal{A}$, $\mathcal{B}$ are unchanged, in $N(n)$ the replacements $A \to A(S_z^k)$ and $\Delta \to \Delta(S_z^k)$ have to be made.

Once $\rho$ is known, $S_z^k$ can be determined self-consistently from
\begin{align*}
	S_z^k = (M - 1) D_0 - \frac{M-1}{M} \frac{2 \kappa}{\Gamma_1} \erw{n}~,
\end{align*} 
with $D_0 = \frac{\Gamma_\uparrow - \Gamma_\downarrow}{\Gamma_\uparrow + \Gamma_\downarrow}$ being the stationary atom polarization.

\begin{figure}
	\centering
	\ifFigureFolder
		\resizebox{!}{.285\textwidth}{\Large \input{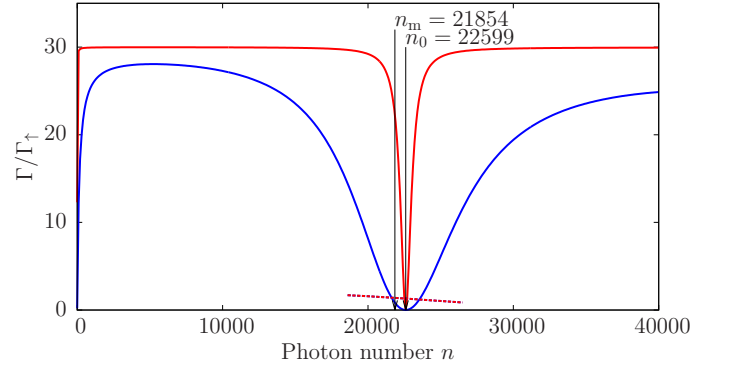}}
	\else 
		\resizebox{!}{.285\textwidth}{\Large \input{Fig3}}
	\fi
	\caption{(Color online) Comparison of the transition rates from $\ket{\uparrow}$ to $\ket{\downarrow}$ creating $1$ photon (solid lines) or $2$ photons (dashed lines), respectively. Due to $\Gamma_\uparrow = \Gamma_\varphi^*$, the 2-photon rates coincide if they are plotted normalized to $\Gamma_\uparrow$. The plot parameters for the red lines are identical to the single-atom case in Fig.~\ref{fig:PhotonStatisticsAndMatrixElement}. The blue lines represent the single-atom case with $10$ times larger pumping and dephasing rates, $\Gamma_\uparrow = .06$, $\Gamma_\downarrow = .001$, $\Gamma_\varphi^* = .01$. The suppression of the $2$-photon rate for small $\theta$ and small pumping rates is visible. Due to our approximations, the plotted $2$-photon rates are only valid near $n_0$.}
	\label{fig:HigherOrderRates}
\end{figure}

\section{Higher order rates}
In the previous discussion, we focused only on energy-conserving transitions in the Hamiltonian. However, the matrix elements $\bra{n}  A \ket{n+m}$ are in fact nonzero for $m \neq 1$. But we will show that there is a range of la\-sing parameters where energy-nonconserving processes are suppressed. 

Energy-nonconserving transitions might drive the system across the squeezing-point $n_0$ where $\bra{n_0} A \ket{n_0+1} = 0$. Therefore, we want the corresponding transition rates to be small at $n_0$. Near $n_0$, we can solve the master-equation containing the energy-nonconserving 2-photon rates $\bra{n} A \ket{n+2}$ while the 1-photon rates $\bra{n} A \ket{n+1}$ vanish. The rate for a 2-photon transition near $n_0$ is
\begin{align*}
	\Gamma_{p \to p+2} &= \frac{M \cdot \mathcal{A}}{1 + \mathcal{B} \cdot \bar{N}(p)} \frac{\abs{\bra{p} A(S_z^k) \ket{p+2}}^2}{(\hbar g)^2} \Gamma_\uparrow~, 
	\\
	\bar{N}(p) &= \frac{(\Delta(S_z^k) - \omega)^2}{4 g^2} \frac{\Gamma_1}{\frac{\Gamma_1}{2} + \Gamma_\varphi^*} + \frac{\abs{\bra{p} A(S_z^k) \ket{p+2}}^2}{(\hbar g)^2}~. 
\end{align*} 
If the master-equation is solved taking into account only the energy-conserving transitions, the corresponding formula for the 1-photon rate is
\begin{align*}
	\Gamma_{p \to p+1} = \frac{M \cdot \mathcal{A}}{1 + \mathcal{B} \cdot N(p)} \frac{\abs{\bra{p} A \ket{p+1}}^2}{\hbar^2 g^2} \Gamma_\uparrow~,
\end{align*}
where $N(p)$ is defined in Eq.~\ref{eqn:NofP}. We now try to modify the lasing parameters $g$ and $\theta$ in order to suppress the 2-photon rate. As we want $n_0$ to be fixed, $p_0$ has to be constant. This reduces the parameter space $(g,\theta)$ to an one-dimensional one, implying $g(\theta) = p_0 \cdot \frac{\omega}{\cos(\theta)}$, and yields the following structure of the transition rates, 
\begin{align*}
	\Gamma_{p \to p+1} 
		&= \frac{p_0^2 \, \omega^2  \tan^2(\theta) \, X_1(p,1)}{1 + \Delta^2 \, X_2 + p_0^2 \, \omega^2 \tan^2(\theta) \, X_3(p,1)} ~,\\
	\Gamma_{p \to p+2} 
		&= \frac{p_0^4 \, \omega^2 \tan^2(\theta) \, X_1(p,2)}{1 + (\Delta - \omega)^2 \, X_2 + p_0^4 \, \omega^2 \tan^2(\theta) \, X_3(p,2)} ~,
\end{align*}
with $X_2 = \left( \frac{\Gamma_1}{2} + \Gamma_\varphi^* \right)^{-2}$ being a constant and $X_1(n,m)$ and $X_3(n,m)$ being functions containing parts of the matrix elements $\bra{n} A \ket{n+m}$. Because of $\omega^2 X_2 \gg 1$, for $\Delta = 0$ the rates are given by
\begin{align*}
	\Gamma_{p \to p+1} 
		&= \frac{p_0^2 \, \omega^2  \tan^2(\theta) \, X_1(p,1)}{1 + p_0^2 \, \omega^2 \tan^2(\theta) \, X_3(p,1)} ~, \\
	\Gamma_{p \to p+2} 
		&= \frac{p_0^4 \tan^2(\theta) \, X_1(p,2)}{X_2 + p_0^4 \tan^2(\theta) \, X_3(p,2)} ~.
\end{align*}
In the limit $\theta \to \pi/2$, both rates are $X_1(p,1)/X_3(p,1) = X_1(p,2)/X_3(p,2) = M \, \Gamma_\uparrow/2$, so there is no suppression. On the other hand, near $n_0$, for each $\omega$ there is a $\theta \to 0$, such that $p_0^2 \, \omega^2 \tan(\theta)^2 X_3(p,1) \ll 1$ and $p_0^4 \tan^2(\theta) X_3(p,2) \ll X_2$. In this limit, we arrive at
\begin{align*}
	\Gamma_{p \to p+1} &= p_0^2 \, \omega^2 \, X_1(p,1) \cdot \theta^2 ~,\\
	\Gamma_{p \to p+2} &= p_0^4 \, \frac{X_1(p,2)}{X_2} \cdot \theta^2 = R(p) \cdot \Gamma_{p \to p+1}~.
\end{align*}
The prefactor $R(p)$ is given by 
\begin{align*}
	R(p) 
	= p_0^2 \left( \frac{\frac{\Gamma_1}{2} + \Gamma_\varphi^*}{\omega} \right)^2 \frac{X_1(p,2)}{X_1(p,1)}~.
\end{align*}
We chose $p_0 = g \cos(\theta)/\omega \ll 1$ fixed and 
\begin{align*}
	\frac{X_1(p,2)}{X_1(p,1)} = \frac{16}{p+2} \left( \frac{L_p^2(4 p_0^2)}{L_p^1(4 p_0^2)} \right)^2
\end{align*}
is a function of $p$ that diverges at $p = n_0$ and fulfills $X_1(p,2)/X_1(p,1) \lesssim 1$ around $n_\mathrm{m}$, where $\rho(n)$ has finite values. So the only way to suppress $R(p)$ is to choose the pumping and dephasing rates small compared to $\omega$. 

Weak pumping decreases the output power of the laser and the 1-photon pumping rate $\Gamma_{p \to p+1}$, which is however necessary for the lasing process. In order to compensate this drop, the number $M$ of atoms has to be enlarged. 

As the suppression relies on the case $\theta \ll \pi/2$, a large $\sigma_z$-coupling to the resonator is needed. Fig.~\ref{fig:HigherOrderRates} illustrates the suppression of the 2-photon rate for $\theta = \pi/16$, comparing two cases whose pumping rates differ by one order of magnitude.

\begin{figure}[t!]
	\centering
	\subfigure[]{
		\tiny
		\newlength{\svgwidth}
		\setlength{\svgwidth}{.35\textwidth}
		\ifFigureFolder
			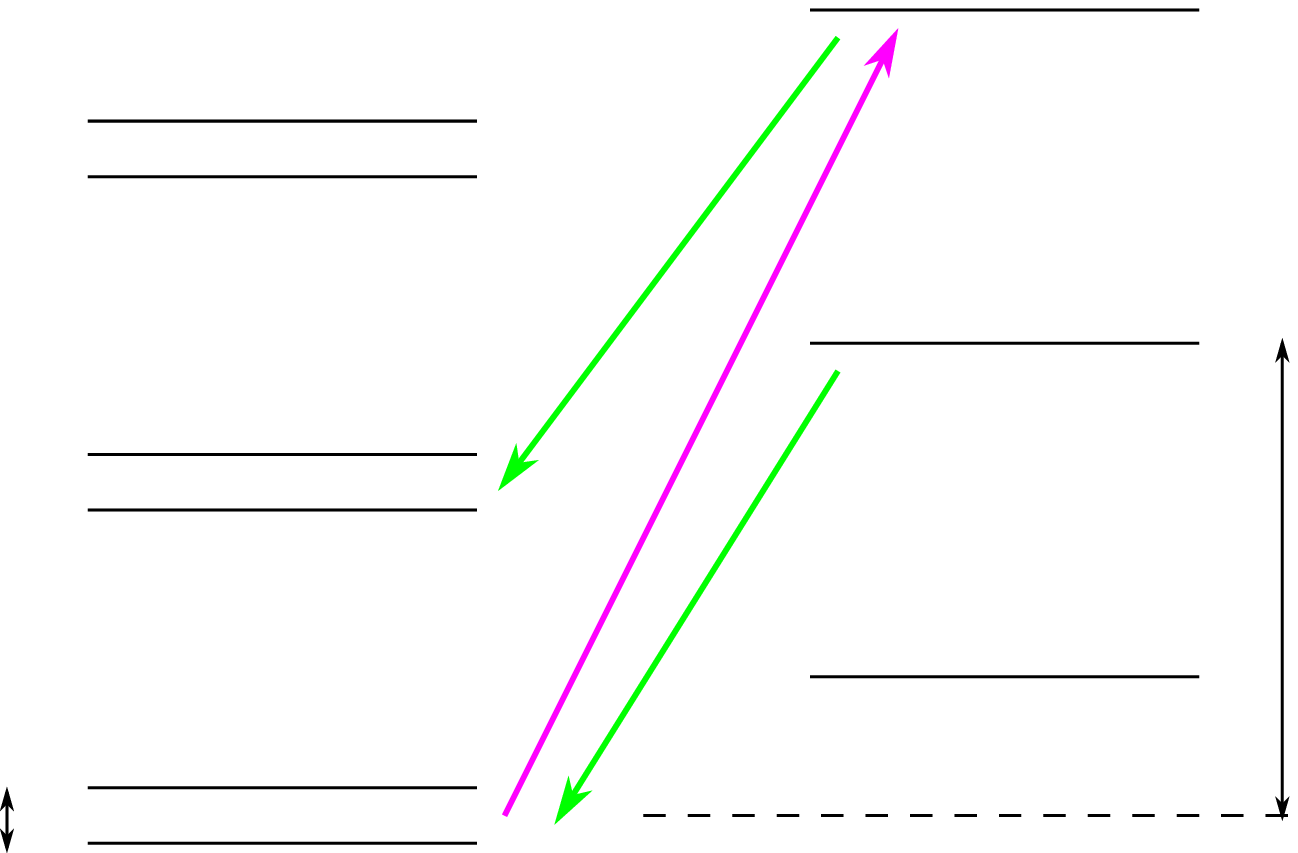		
		\else 
			\input{LevelDiagramEigenbasis.eps_tex}
		\fi
		\label{fig:SpectralFunctionsEigenbasis}
	}
	\subfigure[]{
		\tiny
		\newlength{\svgwidth}
		\setlength{\svgwidth}{.35\textwidth}
		\ifFigureFolder
			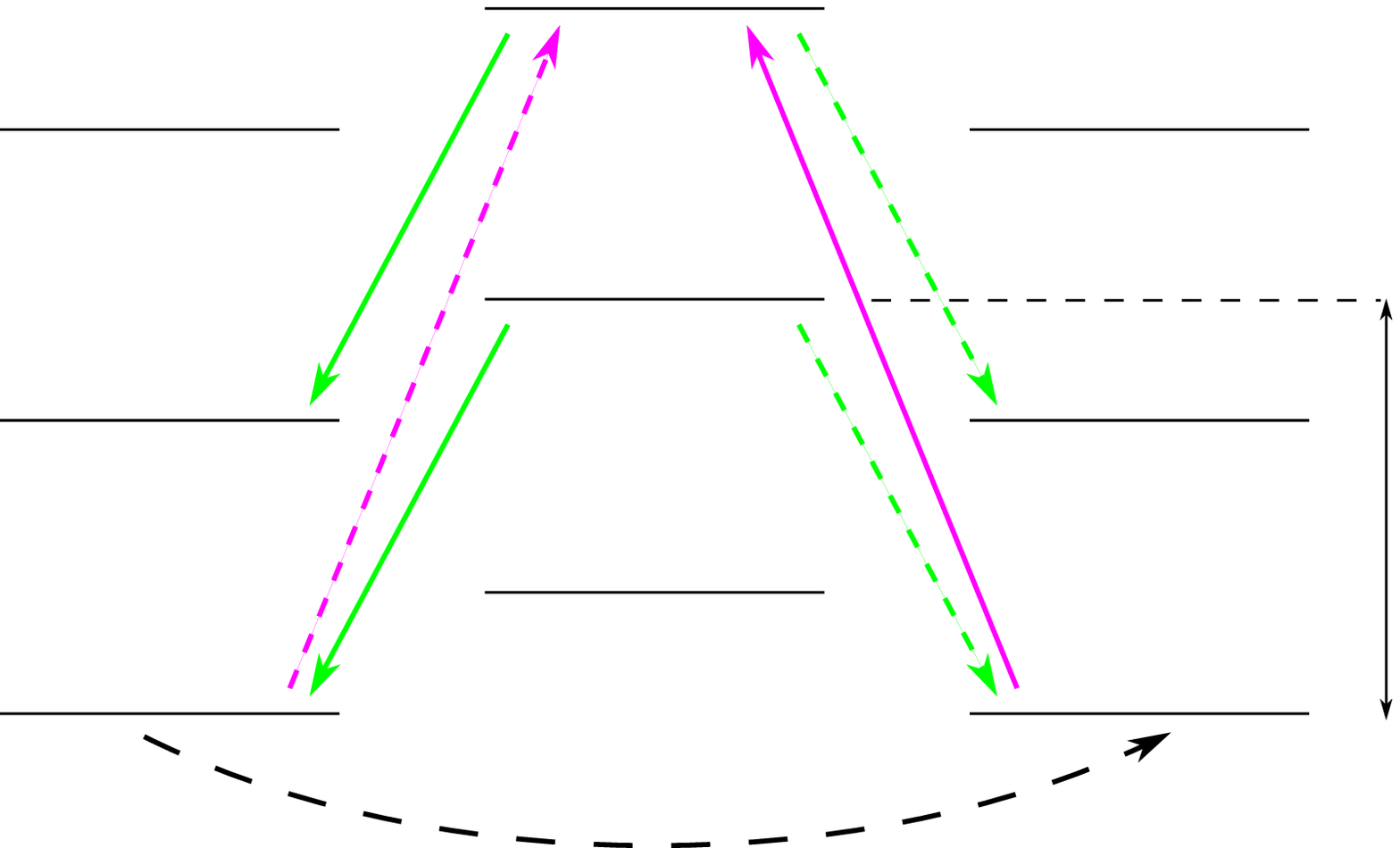
		\else
			\input{LevelDiagramLasingBasis.eps_tex}
		\fi
		\label{fig:SpectralFunctionsLasingBasis}
	}
	\subfigure[]{
		\ifFigureFolder
			\resizebox{!}{.285\textwidth}{\Large \input{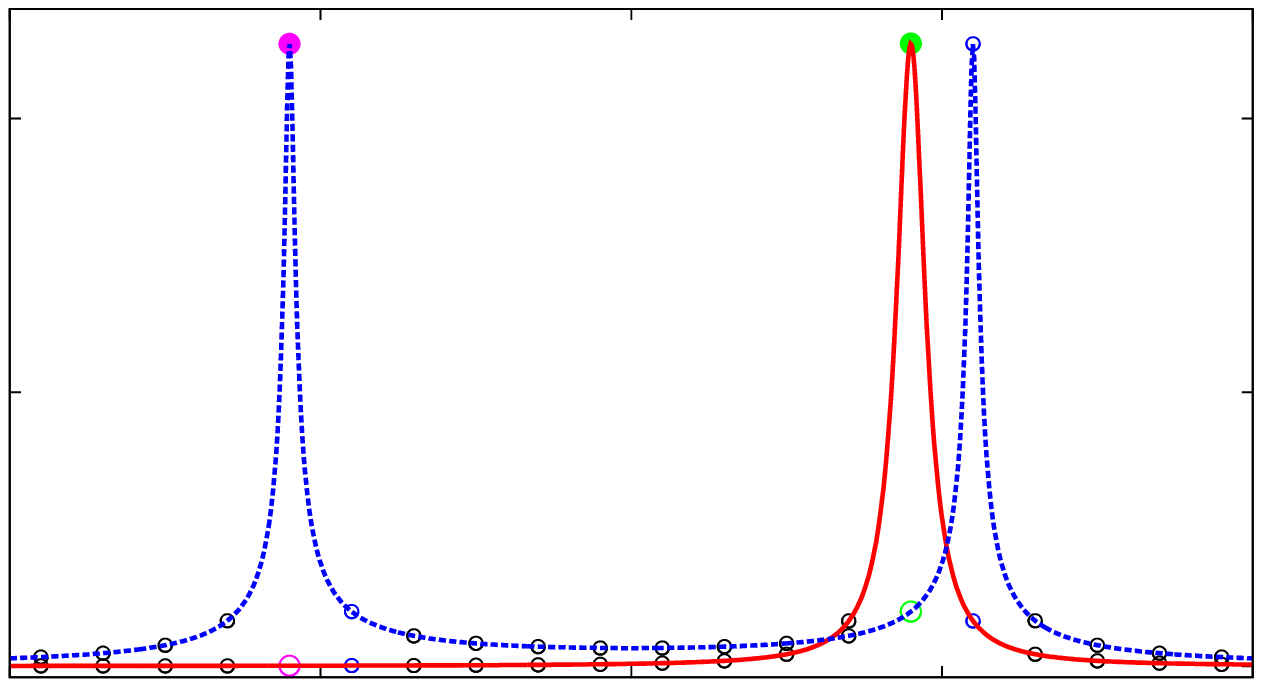}}
		\else			
			\resizebox{!}{.285\textwidth}{\Large \input{Fig4}}
		\fi
		\label{fig:SpectralFunctionsSpectrum}
	}
	\caption{(Color online) (a) Level diagram in the eigenbasis of Eq.~\ref{eqn:HamiltonAfterPolaron}. Green arrows indicate a release of energy into the bath, magenta arrows indicate an absorption out of the bath. (b) Level dia\-gram in the lasing basis. Dashed transitions have to be suppressed by the spectral function of the baths. (c) Spectral functions $S_{1,\uparrow}$ (solid red) and $S_{1,\downarrow}$ (dashed blue) of the baths. $\epsilon > 0$ means release of energy into the bath, $\epsilon < 0$ means absorption out of the bath. Solid and empty circles indicate desired or suppressed transitions, respectively. The transitions of the pumping process are marked in green and magenta. Black transitions correspond to higher order processes. Blue circles indicate the transitions of the inverse pumping process. The parameters of the plot are $\epsilon_1 = 5 \, \omega$, $\gamma' = .28 \, \omega$, $\gamma = .08 \, \omega$.}
	\label{fig:SpectralFunctions}
\end{figure}

\section{Pumping process}
Finally, we show that a pumping process can be implemented and described by a Lindblad-term $L_\mathrm{Q} \rho$, containing effective pumping rates $\Gamma_\uparrow$ and $\Gamma_\downarrow$. 

We model the pumping process by two external reservoirs providing the energy for transitions from the lower lasing state $\ket{\downarrow,n} \equiv \ket{\downarrow} \ket{n}$ to an intermediate state $\ket{1,n}$ (at energy $\epsilon_1$) and from there to the upper lasing state $\ket{\uparrow,n}$, respectively. 
We assume a linear coupling between the reservoirs and the system: $O_{1,\downarrow} X_{1,\downarrow}$ and $O_{1,\uparrow} X_{1,\uparrow}$, respectively. The independent reservoir-operators are denoted by $X_{1,\downarrow}$ and $X_{1,\uparrow}$, respectively. After a polaron transformation, the system-operators are
\begin{align*}
	O_{1,\downarrow} &= \ket{1} \bra{\downarrow} e^{- i p} + \ket{\downarrow} \bra{1} e^{i p}~, \\
	O_{1,\uparrow} &= \ket{1} \bra{\uparrow} e^{i p} + \ket{\uparrow} \bra{1} e^{-i p}~.
\end{align*} 

Due to the factors $e^{\pm i p}$, matrix elements of $O_{1,\uparrow/\downarrow}$ creating more than one photon are nonzero. These multi-photon pumping events disturb the creation of squeezed light and must be suppressed. This suppression is realized by choosing appropriate spectral functions for the baths. 

In order to solve the master-equation, we switch to the eigenbasis of the Hamiltonian~\eqref{eqn:HamiltonAfterPolaron}, which is the usual dressed-state basis of the Jaynes-Cummings model supplemented by the intermediate pumping states $\ket{1,n}$:
\begin{align*}
	\ket{+,n} &= \cos \frac{\chi(n)}{2} \ket{\uparrow,n} + \sin \frac{\chi(n)}{2} \ket{\downarrow,n+1} ~,\\
	\ket{-,n} &= \sin \frac{\chi(n)}{2} \ket{\uparrow,n} - \cos \frac{\chi(n)}{2} \ket{\downarrow,n+1} ~,\\
	\tan \chi(n) &= \frac{2}{\hbar \Delta} \bra{n} A \ket{n+1} ~.
\end{align*} 
For $\theta \ll \pi/4$, higher order matrix elements give small corrections to $\ket{\pm,n}$ which can be calculated perturbatively. The spectral functions of the reservoirs are defined as
\begin{align*}
	S_{1,\uparrow/\downarrow}(\omega) &= \int_{-\infty}^\infty \mathrm{d} \tau \, e^{- i \omega \tau} \erw{X_{1,\uparrow/\downarrow}(\tau) X_{1,\uparrow/\downarrow}(0)} ~.
\end{align*}
The diagonalized Hamiltonian is
\begin{align*}
	H &= \frac{1}{2} \sum_{n=0}^\infty E(n) \left[ \ket{+,n} \bra{+,n} - \ket{-,n} \bra{-,n} \right] \\
	&+ \sum_{n=0}^\infty \sum_{s = \pm} \hbar \omega \left( n + \frac{1}{2} \right) \ket{s,n} \bra{s,n} \\
	&- \frac{\epsilon}{2} \ket{\downarrow,0} \bra{\downarrow,0} + \sum_{n=0}^\infty \left( \epsilon_1 + n \hbar \omega \right) \ket{1,n} \bra{1,n} ~,\\
	E(n) &= \sqrt{\hbar^2 \Delta^2 + 4 (\bra{n} A \ket{n+1})^2} ~.
\end{align*} 
Fig.~\ref{fig:SpectralFunctionsEigenbasis} shows a sketch of the level diagram including the pumping process that increases the number of photons in each step by one,
\begin{align*}
	\ket{1,n} \to \ket{\pm,n} \to \ket{1,n+1}~.
\end{align*} 
We assume $E(n) \ll \epsilon_1, \hbar \omega$. Choosing $\epsilon_1 \gg 0$, the system relaxes from $\ket{1,n}$ to $\ket{\pm,n}$ releasing the energy $\epsilon_1 - \frac{\hbar \omega}{2}$ to the baths (green arrows). From there, a transition is made to $\ket{1,n+1}$ taking the energy $\epsilon_1 + \frac{\hbar \omega}{2}$ out of the baths (magenta arrows). Higher order processes take or release energies which differ from $\epsilon_1 \pm \frac{\hbar \omega}{2}$ by an integer multiple of $\hbar \omega$. Therefore, the spectral functions $S_{1,\uparrow}$ and $S_{1,\downarrow}$ of the baths must be peaked at $\epsilon_1 \mp \frac{\hbar \omega}{2}$, and the peaks must be sharp enough to give small values at all other energies differing by an integer multiple of $\hbar \omega$.

If $E(n) \ll \hbar \omega$, we can perform a similar analysis in the $\ket{\uparrow/\downarrow,n}$-basis. The corresponding level diagram is sketched in Fig.~\ref{fig:SpectralFunctionsLasingBasis}. In order to arrive at a directed pumping process that increases the number of photons in each step, the dashed transitions have to be suppressed. Therefore, the spectral function $S_{1,\uparrow}$ and $S_{1,\downarrow}$ must be chosen differently.

Fig.~\ref{fig:SpectralFunctionsSpectrum} shows, that these conditions can be fulfilled assuming that the pumping bath is a pumping laser with the spectral function of a harmonic oscillator and the relaxation bath is some dissipation process with a Lorentzian spectral function,
\begin{align*}
	S_{1,\downarrow}(\epsilon) &= \frac{S_0}{\sqrt{(\epsilon^2 - \omega_\mathrm{r}^2)^2 + 4 \gamma^2 \epsilon^2}}~, \\
	S_{1,\uparrow}(\epsilon) &= \frac{S_0'}{\pi} \frac{\gamma'}{(\epsilon - \omega_\mathrm{r})^2 + {\gamma'}^2}~.
\end{align*}
$\omega_\mathrm{r}$ denotes the resonance frequencies, which differ for $S_{1,\uparrow}$ and $S_{1,\downarrow}$ by $\hbar \omega$. $\gamma$ is the damping parameter of the oscillator, $\gamma'$ the width parameter of the Lorentzian function and $S_0$ and $S_0'$ are some constants. Hence, transitions that don't belong to the pumping process can be suppressed and effective rates for the transitions $\ket{\downarrow,n} \leftrightarrow \ket{\uparrow,n}$ can be calculated and written as pumping rates into a Lindblad-term. In general, the pumping rates $\Gamma_\downarrow$ and $\Gamma_\uparrow$ depend weakly on $n$. 

\section{Fano factor in the multi-atom case}
In the multi-atom case, the polaron transformation is generalized to the one defined in Eq.~\ref{eqn:PolaronMAtoms}. Therefore, the corrections to the Fano factor due to the polaron transformation change. They give a constraint on the maximum number of atoms if a certain Fano factor should be reached. The correction to the photon number operator due to the polaron transformation is 
\begin{align*}
	a^\dagger_\mathrm{p} a_\mathrm{p} &= a^\dagger a - \xi (a + a^\dagger) + \xi^2 ~,\\
	\xi &= \sum_{i=1}^M p_0 \sigma_z^i ~.
\end{align*}
Therefore, the nominator of $F$ is given by
\begin{align*}
	\erw{n^2}_\mathrm{c} - \erw{n}^2_\mathrm{c} 
	&= \erw{n^2} - \erw{n}^2 \\
	&- \erw{\xi \left( a^\dagger a (a + a^\dagger) + (a + a^\dagger) a^\dagger a \right)} \\
	&+ 2 \erw{a^\dagger a} \erw{\xi (a + a^\dagger)} \\
	&+ \erw{\xi^2 \left( a^2 + (a^\dagger)^2 + 4 a^\dagger a + 1 \right)} \\
	&- \erw{\xi (a + a^\dagger)}^2 - 2 \erw{a^\dagger a} \erw{\xi^2} \\
	&- 2 \erw{\xi^3 (a + a^\dagger)} + 2 \erw{\xi}^2 \erw{\xi (a + a^\dagger)} \\
	&+ \erw{\xi^4} - \erw{\xi^2}^2~,
\end{align*}
where the subscript $\mathrm{c}$ denotes the cavity-frame, as introduced above. Most of the terms cancel due to the following reasons:
\begin{enumerate}
	\item The multi-atom calculation is performed in a mean-field approximation, therefore $\erw{\xi^r} = \erw{\xi}^r$ and $\erw{\xi^r \xi a^{(\dagger)}} = \erw{\xi}^r \erw{\xi a^{(\dagger)}}$ for $r \in \mathbb{N}$. 
	\item $\erw{\xi a^{(\dagger)}} = 0$ as these terms are proportional to the energy-nonconserving matrix elements $\rho_{\sigma \, n, \sigma \, n \pm 1} = 0$, with $\sigma \in \{ \uparrow, \downarrow \}$.
\end{enumerate}
The remaining non-vanishing terms are
\begin{align*}
	\erw{n^2}_\mathrm{c} - \erw{n}^2_\mathrm{c} 
	&= \erw{n^2} - \erw{n}^2 \\
	&+ 4 \erw{\xi^2 a^\dagger a} + \erw{\xi^2} - 2 \erw{a^\dagger a} \erw{\xi^2}~.
\end{align*}
The expectation values yield in the limit $\erw{n} \gg 1$:
\begin{align*}
	\erw{\xi^2} &= p_0^2 M + p_0^2 \frac{M}{M-1} \left( S_z^k \right)^2 ~,\\
	\erw{\xi^2 a^\dagger a} &= \erw{n} \erw{\xi^2} - \erw{n} p_0^2 \frac{2 \kappa}{\Gamma_1} S_z^k F_0 ~,
\end{align*}
where $S_z^k$ is the self-consistent atomic polarization for an arbitrarily chosen atom $k$ and $F_0 = (\erw{n^2} - \erw{n}^2)/\erw{n}$ is the Fano factor of the photon-distribution in the polaron-frame. In the limit $\erw{n} \gg \erw{\xi^2}$, the Fano factor is
\begin{align*}
	F &\approx F_0 \left( 1 - p_0^2 \frac{8 \kappa}{\Gamma_1} S_z^k \right) + 2 \erw{\xi^2}~.
\end{align*}
If the laser produces classical light, $S_z^k \approx 0$ because the photon number in the cavity is defined by the balance of pumping and loss rates. On the other hand, if the laser produces squeezed light, the number of photons is defined by the root of the coupling matrix element and $S_z^k \lesssim M$. Therefore, we write $S_z^k = \eta \cdot M$ with $\eta \in [0,1]$. For typical lasing parameters $p_0$ and $M = \mathcal{O}(100)$ we have $p_0^2 M \ll 1$, but $p_0^2 M^2$ is of the order of unity. Therefore, the corrections to the Fano factor due to the polaron transformation can be written as
\begin{align*}
	F \approx F_0 + 2 p_0^2 \eta^2 M^2~.
\end{align*}
The second term of $F$ can be arbitrarily large for large $M$. If the Fano factor should be smaller than a certain threshold $F_\mathrm{max}$ and if $p_0$ is fixed, we arrive at the constraint 
\begin{align*}
	M \leq \sqrt{\frac{F_\mathrm{max} - F_0}{2 p_0^2}}~.
\end{align*}
Decreasing $p_0$ shifts the roots of the coupling matrix element to higher photon numbers and weakens this constraint.

\section{Conclusion}
In this paper we showed, that artificial atoms with both $\sigma_x$- and $\sigma_z$-coupling offer a way to construct a laser that produces photon-number squeezed light. The maxi\-mum photon number can be modified via the coupling strength $g$ and the mixing angle $\theta$. In contrast to other proposals, usual coupling strengh of the order of $10^{-3} \omega$ give rise to large photon numbers of the order of $10^4$. Once a maximum photon number is chosen, $g$ and $\theta$ can be adjusted to suppress energy-nonconserving transitions that would otherwise destroy the squeezed state. Furthermore, a pumping process can be implemented using two external baths. 

Coupling multiple artificial atoms to a common reso\-na\-tor has already been demonstrated experimentally by the construction of a quantum metamaterial consisting of $20$ superconducting flux qubits \cite{Metamaterial_Pascal}. The individual qubits exhibited a mixing angle $\theta \approx 1.18$ and a bare coupling strength $g/\omega \approx 5 \cdot 10^{-5}$ to the third resonator mode at $\omega_3/(2 \pi) = 3 \cdot 2.594\,\mathrm{GHz}$. In this setup, $g/\omega$ is actually quite small. A larger, but still realistic, coupling strength of $g/\omega = 4 \cdot 10^{-3}$ and a typical resonator decay rate of $\kappa/\omega = 10^{-5}$ yield for $M=200$ atoms and rates of $\Gamma_\uparrow/\omega = 0.05$, $\Gamma_\downarrow/\omega = 0.0001$ and $\Gamma_\varphi^*/\omega = 0.001$ an average photon number of $\erw{n} \approx 381 700$ and a Fano factor $F \approx 0.08$. Hence, using modified qubits with a smaller mixing angle $\theta$, an experimental realization of the presented laser is possible and promising with current qubit technology. 

\section{Acknowledgement}
We acknowledge helpful discussions with J. Lepp\"akangas and G. Sch\"on and funding from the DFG-grant MA 6334/3-1. 

\newpage

\section*{Appendix}
\begin{widetext}
The $4 \times 4$-matrix $M$ and the vector $\vec{A}$ defined above in the calculation of $\rho$ have the following form ($\hbar=1$):
\begin{align*}
	M &= \begin{pmatrix}
		- i \omega (p - q) - \Gamma_\downarrow - \Gamma_\uparrow & i \bra{q+1} A^\dagger \ket{q} & - i \bra{p} A \ket{p+1} & 0 \\
		i \bra{q} A \ket{q+1} & - i \Delta - i \omega(p - q) - \frac{\Gamma_\downarrow}{2} - \frac{\Gamma_\uparrow}{2} - \Gamma_\varphi^* & 0 & - i \bra{p} A \ket{p+1} \\
		- i \bra{p+1} A^\dagger \ket{p} & 0 & i \Delta - i \omega (p-q) - \frac{\Gamma_\downarrow}{2} - \frac{\Gamma_\uparrow}{2} - \Gamma_\varphi^* & i \bra{q + 1} A^\dagger \ket{q} \\
		0 & - i \bra{p+1} A^\dagger \ket{p} & i \bra{q} A \ket{q+1} & - i \omega (p-q) - \Gamma_\uparrow - \Gamma_\downarrow
	\end{pmatrix} \\
	\vec{A}(\rho_\mathrm{r}) &= \begin{pmatrix}
		\Gamma_\uparrow \cdot (\rho_\mathrm{r})_{p,q} \\
		0 \\
		0 \\
		\Gamma_\downarrow \cdot (\rho_\mathrm{r})_{p+1,q+1}
	\end{pmatrix}
\end{align*} 
\end{widetext}

\end{document}